\documentclass{elsart3-1}

\usepackage[english,francais]{babel}

\usepackage{url,graphicx}
\usepackage[dvips]{epsfig}
\usepackage{latexsym,color}
\usepackage{amscd,amssymb,verbatim}
\newcommand{\be}{\begin{enumerate}}
\newcommand{\ee}{\end{enumerate}}

\newcommand{\bo}{\partial}
\newcommand{\atom}{\textrm{\tt Atom}\,}

\newcommand{\ca}{{\mathcal A}}
\newcommand{\cc}{{\mathcal C}}
\newcommand{\car}{{\mathcal R}}
\newcommand{\cd}{{\mathcal D}}
\newcommand{\cm}{{\mathcal M}}

\newcommand{\cu}{{\mathcal U}}

\newcommand{\dz}{{\mathbb Z}}

\newcommand{\id}{{\textrm{id}}}

\newcommand{\om}{\Omega}

\newcommand{\fk}{{\mathfrak K}}

\newcommand{\lra}{\longrightarrow}

\newcommand{\nin}{\noindent}
\newcommand{\pr}{\noindent{\bf Proof. }}
\newcommand{\sm}{\setminus}

\newcommand{\wti}{\widetilde}

\newcommand{\ti}{\tilde}

\newtheorem{theorem}{Theorem}[section]

\newtheorem{e-proposition}[theorem]{Proposition}

\newtheorem{e-definition}[theorem]{Definition\rm}
\newtheorem{remark}{\it Remark\/}

\newtheorem{theoreme}{Th\'eor\`eme}[section]

\newtheorem{definition}[theoreme]{D\'efinition\rm}

\def\og{\leavevmode\raise.3ex\hbox{$\scriptscriptstyle\langle\!\langle$~}}
\def\fg{\leavevmode\raise.3ex\hbox{~$\!\scriptscriptstyle\,\rangle\!\rangle$}}

\begin{document}

\begin{frontmatter}
\selectlanguage{english}

\title{Discrete Morse Theory for free chain complexes}

\vspace{-2.6cm}

\selectlanguage{francais}
\title{Th\'eorie de Morse pour des complexes de chaines libres}

\selectlanguage{english}

\author{Dmitry N. Kozlov}
\ead{dkozlov@inf.ethz.ch}
\ead[url]{http://www.ti.inf.ethz.ch/people/kozlov.html}

\address{Eidgen\"ossische Technische Hochschule, Z\"urich, Switzerland}
\thanks{Research supported by Swiss National Science Foundation Grant PP002-102738/1}

\begin{abstract}
 
We extend the combinatorial Morse complex construction to the
arbitrary free chain complexes, and give a~short, self-contained, and
elementary proof of the quasi-isomorphism between the original chain
complex and its Morse complex.

Even stronger, the main result states that, if $C_*$ is a free chain
complex, and $\cm$ an acyclic matching, then $C_*=C_*^\cm\oplus T_*$,
where $C_*^\cm$ is the Morse complex generated by the critical
elements, and $T_*$ is an acyclic complex. {\it To cite this article:
D.N.\ Kozlov, C.\ R.\ Acad.\ Sci.\ Paris, Ser.\ I ??? (2005).}

\vskip 0.5\baselineskip

\selectlanguage{francais}
\noindent{\bf R\'esum\'e}
\vskip 0.5\baselineskip
\noindent
On \'etend la construction du complex de Morse combinatoire aux complexes 
de chaines libres g\'en\'erals , et on donne une demonstration br\`eve et \'el\'ementaire du quasi-isomorphisme  entre le complex de chaines original et son complex de Morse.

Plus profondement, le r\'esultat principal dit que, si $C_*$ est un complex de chaines libres, et $\cm$ est une correspondence acyclique, puis $C_*=C_*^\cm\oplus T_*$,
et $C_*^\cm$ est le complexe de Morse g\'en\'er\'e par les \'el\'ements critiques, et $T_*$ est un complex acyclique. {\it Pour citer cet article~: D.N.\ Kozlov, C.\ R.\ Acad.\ Sci.\ Paris, Ser.\ I ??? (2005).}

\end{abstract}
\end{frontmatter}




\selectlanguage{francais}
\section*{Version fran\c{c}aise abr\'eg\'ee}

La th\'eorie de Morse discr\`ete a \'et\'e introduite par
Forman~\cite{Fo}, et elle a prouv\'e d'\^{e}tre profitable pour des
calculations en combinatoire topologique. Il a \'{e}t\'{e} demontr\'e
en \cite[Theorem 8.2]{Fo} que, donn\'e une fonction de Morse
discr\`ete~\cite[Definition 2.1]{Fo} sur un complex CW fini $K$, le
complex de chaines cellulair $C_*(K;\dz)$ est quasi-isomorphique au
complex de Morse combinatoire associ\'e.

Dans cet article, on \'etend cette construction au cas de complexes de
chaines libres g\'e\'nerals. On pr\'esente une demonstration
independente et simple dans cette g\'e\'neralit\'e, en particulier, on
arrive a une demonstration nouvelle et \'el\'ementaire des r\'esultats
de Forman. Sur un niveau plus \'el\'ev\'e, on peut regarder notre
demonstration comme un analogue algebrique des arguments pr\'esent\'es
en \cite[Theorem 3.2]{Ko}.

Soit $\car$ un anneau commutative g\'en\'eral avec un element neutre. 
On dit qu' un complex de chaines
$C_*$ aux $\car$-modules
$\dots\stackrel{\bo_{n+2}}{\lra} C_{n+1}\stackrel{\bo_{n+1}}{\lra} 
C_{n}\stackrel{\bo_{n}}{\lra}
C_{n-1}\stackrel{\bo_{n-1}}{\lra}\dots$, est {\it libre} si 
chaque $C_n$ est un $\car$-module libre g\'en\'er\'e au fini. 
Si les indices sont clairs, on ecrit $\bo$ au lieux de $\bo_{n}$
On  demande que $C_*$ est born\'e a droite.

Soup\c{c}onn\'e qu'on a choisit un base (un ensemble de g\'en\'erateurs libres)
$\Omega_n$ pour chaque $C_n$. En ce cas, on dit qu'on a choisit un base
$\Omega=\bigcup_{n}\Omega_n$ pour $C_*$, et on ecrit 
$(C_*,\Omega)$ pour un complex de chaines avec un base. 
Un complex de chaines libre avec un base est le charact\`ere principal  
de cet article.

\begin{definition} \label{fdef:pm}
Soit $(C_*,\Omega)$ un complex de chaines libre avec un base. 
\begin{enumerate}
\item [(1)] Une {\bf correspondence partielle} $\cm\subseteq\Omega\times\Omega$ sur $(C_*,\Omega)$
est une correspondence partielle sur le diagramme Hasse de $P(C_*,\Omega)$,
tel que, si $b\succ a$, et $b$ et $a$ sont en correspondence, i.e.\ si $(a,b)\in
M$, donc $w(b\succ a)$ est invertible.

\item [(2)] Une correspondence partielle sur  $(C_*,\Omega)$ est {\bf acyclique}, 
s'il n'y a pas de cycle 
\begin{equation} \label{feq:cycle}
 d(b_1)\prec b_1\succ d(b_2)\prec b_2\succ d(b_3)\prec\dots
\succ d(b_n)\prec b_n\succ d(b_1),
\end{equation}
avec $n\geq 2$, est tous $b_i\in\cu(\Omega)$ diff\'erents.
\end{enumerate}
\end{definition}

\begin{definition}
Soit $(C_*,\Omega)$ un complex de chaines libre avec un base, et soit $\cm$
une correspondence acyclique.  Le {\bf complex de Morse}
$\dots\stackrel{\bo^\cm_{n+2}}{\lra} C_{n+1}^\cm \stackrel{\bo^\cm_{n+1}}{\lra} C_{n}^\cm\stackrel{\bo^\cm_n}{\lra} C_{n-1}^\cm \stackrel{\bo^\cm_{n-1}}{\lra}\dots$,
est defini comme suit. Le $\car$-module $C_{n}^\cm$ est librement g\'en\'er\'e
par les elements de $\cc_n(\Omega)$. L'operateur de borne est defini par 
$\bo^\cm_n(s)=\sum_{p} w(p)\cdot p_\bullet$,
pour $s\in\cc_n(\Omega)$, quand la somme est prise sur tous les chemins 
altern\'es $p$ qui satisfaient $p^\bullet=s$.
\end{definition}

Le complex de chaines $\dots\lra 0\lra\car\stackrel\id\lra\car\lra
0\lra\dots$, dans lequels les seules modules non-trivials se trouvent 
en dimension $d$ et $d-1$, on l'appelle un~{\it complex de chaines atomique}, 
et on ecrit $\atom(d)$.

Le r\'esultat principal de cet article est comme suit:

\begin{theoreme}
Soup\c{c}onne qu'on a un complex de chaines libre avec un base
$(C_*,\Omega)$,
et une correspondence acyclique $\cm$. Puis, $C_*$ se decompose en un somme direct de complexes de chaines $C^\cm_*\oplus T_*$, et  $T_*\simeq\bigoplus_{(a,b)\in\cm}\atom({\dim b})$.
\end{theoreme}

\selectlanguage{english}
\section{Acyclic matchings on chain complexes and the Morse complex.}


Discrete Morse theory was introduced by Forman, see \cite{Fo}, and it
proved to be useful in various computations in topological
combinatorics. It was shown, \cite[Theorem 8.2]{Fo}, that, given a
discrete Morse function, \cite[Definition 2.1]{Fo}, on a~finite CW
complex $K$, the cellular chain complex $C_*(K;\dz)$ is
quasi-isomorphic to the associated combinatorial Morse complex.

In this paper, we extend this construction to the case of arbitrary
free chain complexes. We give an independent, simple, and
self-contained proof in this generality, in particular furnishing
a~new elementary and short derivation of the Forman's result. On a~higher level, our proof can be viewed as an algebraic analog of the argument given in \cite[Theorem 3.2]{Ko}.

Let $\car$ be an arbitrary commutative ring with a~unit. We say that
a~chain complex $C_*$ consisting of $\car$-modules
$\dots\stackrel{\bo_{n+2}}{\lra} C_{n+1}\stackrel{\bo_{n+1}}{\lra} 
C_{n}\stackrel{\bo_{n}}{\lra}
C_{n-1}\stackrel{\bo_{n-1}}{\lra}\dots$, is {\it free} if each $C_n$
is a~finitely generated free $\car$-module. When the indexing is
clear, we simply write $\bo$ instead of $\bo_{n}$. We require $C_*$ to
be bounded on the right.

Assume that we have chosen a basis (i.e.\ a~set of free generators)
$\Omega_n$ for each $C_n$. In this case we say that we have chosen
a~basis $\Omega=\bigcup_{n}\Omega_n$ for $C_*$, and we write
$(C_*,\Omega)$ to denote a~chain complex with a~basis. A~free chain
complex with a~basis is the main character of this paper.

Given a~free chain complex with a~basis $(C_*,\Omega)$, and two
elements $\alpha\in C_n$, and $b\in\Omega_n$, we denote the
coefficient of $b$ in the representation of $\alpha$ as a~linear
combination of the elements of $\Omega_n$ by $\fk_\Omega(\alpha,b)$,
or, if the basis is clear, simply by~$\fk(\alpha,b)$. For $x\in C_n$
we write $\dim x=n$. By convention, we set $\fk_\Omega(\alpha,b)=0$ if
the dimensions do not match, i.e., if $\dim\alpha\neq\dim b$.

Note that a~free chain complex with a~basis $(C_*,\Omega)$ can be
represented as a~ranked poset $P(C_*,\Omega)$, with $\car$-weights on
the order relations.  The elements of rank $n$ correspond to the
elements of $\Omega_n$, and the weight of the covering relation
$b\succ a$, for $b\in\Omega_n, a\in\Omega_{n-1}$, is simply defined by
$w_\Omega(b\succ a) :=\fk_\Omega(\bo b,a)$. In other words, $\bo
b=\sum_{b\succ a} w_\Omega(b\succ a) a$, for each $b\in
\Omega_n$. Again, if the basis is clear, we simply write~$w(b\succ
a)$.

\begin{e-definition} \label{def:pm}
Let $(C_*,\Omega)$ be a free chain complex with a~basis.  A {\bf
partial matching} $\cm\subseteq\Omega\times\Omega$ on $(C_*,\Omega)$
is a partial matching on the covering graph of $P(C_*,\Omega)$, such
that if $b\succ a$, and $b$ and $a$ are matched, i.e.\ if $(a,b)\in M$,
then $w(b\succ a)$ is invertible.
\end{e-definition}

\begin{remark}
Note that the Definition \ref{def:pm} is different from
\cite[Definition 2.1]{Fo}. The latter is a topological definition, and
has the condition that the matched cells form a~regular pair (in the
CW sense). In our algebraic setting it suffices to require the
invertibility of the covering weight.
\end{remark}

Given such a~partial matching $\cm$, we write $b=u(a)$, and $a=d(b)$,
if $(a,b)\in\cm$. We denote by $\cu_n(\Omega)$ the set of all
$b\in\Omega_n$, such that $b$ is matched with some $a\in\Omega_{n-1}$,
and, analogously, we denote by $\cd_n(\Omega)$ the set of all
$a\in\Omega_{n}$, which are matched with some $b\in\Omega_{n+1}$.  We
set $\cc_n(\Omega):=\Omega_n\sm\{\cu_n(\Omega)\cup\cd_n(\Omega)\}$ to
be the set of all unmatched basis elements; these elements are called
{\it critical}.  Finally, we set
$\cu(\Omega):=\bigcup_{n}\cu_n(\Omega)$,
$\cd(\Omega):=\bigcup_{n}\cd_n(\Omega)$, and
$\cc(\Omega):=\bigcup_{n}\cc_n(\Omega)$.

Given two basis elements $s\in\Omega_n$ and $t\in\Omega_{n-1}$, an
{\it alternating path} is a~sequence
\begin{equation} \label{eq:apath}
p=(s\succ d(b_1)\prec b_1\succ d(b_2)\prec b_2\succ\dots\succ d(b_n)\prec b_n \succ t),
\end{equation}
where $n\geq 0$, and all $b_i\in\cu(\Omega)$ are distinct. We use the
notations $p^\bullet=s$ and $p_\bullet=t$. The {\it weight} of
such an~alternating path is defined to be the quotient
\[w(p):=(-1)^n\frac{w(s\succ d(b_1))\cdot w(b_1\succ d(b_2))\cdot\dots \cdot
w(b_n\succ t)}{w(b_1\succ d(b_1))\cdot w(b_2\succ d(b_2))\cdot\ldots\cdot w(b_n\succ d(b_n))}.
\]

\begin{e-definition}
A partial matching on $(C_*,\Omega)$ is called {\bf acyclic}, if there
does not exist a cycle
\begin{equation} \label{eq:cycle}
 d(b_1)\prec b_1\succ d(b_2)\prec b_2\succ d(b_3)\prec\dots
\succ d(b_n)\prec b_n\succ d(b_1),
\end{equation}
with $n\geq 2$, and all $b_i\in\cu(\Omega)$ being distinct.
\end{e-definition}

There is a nice alternative way to reformulate the notion of acyclic
matching.

\begin{e-proposition} \label{prop:linext}
A partial matching on $(C_*,\Omega)$ is acyclic if and only if there
exists a~linear extension of $P(C_*,\Omega)$, such that, in this
extension $u(a)$ follows directly after $a$, for all
$a\in\cd(\Omega)$.

This extension can always be chosen so that, restricted to
$\cd(\Omega)\cup\cc(\Omega)$, it does not decrease rank.
\end{e-proposition}
\pr If such an extension $L$ exists, then following a cycle (\ref{eq:cycle})
left to right we always go down in the order $L$ (more precisely,
moving one position up is followed by moving at least two positions
down), hence a~contradiction.

Assume that the matching is acyclic, and define $L$ inductively. Let
$Q$ denote the set of elements which are already ordered in $L$. We
start with $Q=\emptyset$. Let $W$ denote the set of the lowest rank elements in $P(C_*,\Omega)\sm Q$. At each step we have one of the following cases.

\nin {\it Case 1.} One of the elements $c$ in $W$ is critical. 
Then simply add $c$ to the order $L$ as the largest element, and proceed
with $Q\cup\{c\}$.

\nin {\it Case 2.} All elements in $W$ are matched. The covering graph
induced by $W\cup u(W)$ is acyclic, hence the total number of edges is
at most $2|W|-1$. It follows that there exists $a\in W$, such that
$P(C_*,\Omega)_{<u(a)}\setminus Q=\{a\}$. Hence, we can add elements
$a$ and $u(a)$ on top of $L$ and proceed with $Q\cup\{a,u(a)\}$.
\qed

\begin{e-definition}
Let $(C_*,\Omega)$ be a free chain complex with a~basis, and let $\cm$
be an~acyclic matching.  The {\bf Morse complex}
$\dots\stackrel{\bo^\cm_{n+2}}{\lra} C_{n+1}^\cm \stackrel{\bo^\cm_{n+1}}{\lra} C_{n}^\cm\stackrel{\bo^\cm_n}{\lra} C_{n-1}^\cm \stackrel{\bo^\cm_{n-1}}{\lra}\dots$,
is defined as follows. The $\car$-module $C_{n}^\cm$ is freely
generated by the elements of $\cc_n(\Omega)$. The boundary operator is defined by
$\bo^\cm_n(s)=\sum_{p} w(p)\cdot p_\bullet$,
for $s\in\cc_n(\Omega)$, where the sum is taken over all alternating paths
$p$ satisfying $p^\bullet=s$.
\end{e-definition}

Again, if the indexing is clear, we simply write $\bo^\cm$ instead
of~$\bo^\cm_n$.

Given a free chain complex with a basis $(C_*,\Omega)$, we can choose
a~different basis $\wti\Omega$ by replacing each $a\in\cd_n(\Omega)$ by
$\ti a=w(u(a)\succ a)\cdot a$. Since 
\begin{equation} \label{eq:coeff}
 \fk_{\wti\Omega}(x,\ti a)=\fk_\Omega(x,a)/w(u(a)\succ a),
\end{equation}
for any $x\in\Omega_n$, we see that the weights of those alternating
paths, which do not begin with or end in an element from
$\cd_n(\Omega)$, remain unaltered, as the quotient $w(x\succ
z)/w(y\succ z)$ stays constant as long as $x,y\neq a$. In particular,
the Morse complex will not change. On the other hand,
by~(\ref{eq:coeff}), $w_{\wti\om}(u(a)\succ a)=1$, for all $a\in\cd(\wti\Omega)$, so the total weight
of the alternating path in (\ref{eq:apath}) will simply become
\[w_{\wti\om}(p)=(-1)^n w_{\wti\om}(s\succ d(b_1))\cdot w_{\wti\om}(b_1\succ d(b_2))\cdot
\dots \cdot w_{\wti\om}(b_n\succ t).\]
Because of these observations, we may always replace any given basis
of $C_*$ with the basis $\wti\Omega$ satisfying $w_{\wti\om}(u(a)\succ a)=1$, for all $a\in\cd(\wti\Omega)$.

\section{The main theorem.}

The chain complex $\dots\lra 0\lra\car\stackrel\id\lra\car\lra
0\lra\dots$, where the only nontrivial modules are in the dimensions
$d$ and $d-1$, is called an~{\it atom chain complex}, and is denoted
by $\atom(d)$.

The main result brings to light a~certain structure in $C_*$. Namely,
by choosing a~different basis, we will represent $C_*$ as a~direct sum
of two chain complexes, of which one is a~direct sum of atom chain
complexes, in particular acyclic, and the other one is isomorphic to
$C^\cm_*$. For convenience, the choice of basis will be performed in
several steps, one step for each matched pair of the basis elements.

\begin{theorem}
Assume that we have a free chain complex with a basis $(C_*,\Omega)$,
and an~acyclic matching $\cm$. Then $C_*$ decomposes as a~direct sum
of chain complexes $C^\cm_*\oplus T_*$, where
$T_*\simeq\bigoplus_{(a,b)\in\cm}\atom({\dim b})$.
\end{theorem}

\pr
To start with, let us choose a linear extension $L$ of the partially
ordered set $P(C_*,\Omega)$ satisfying the conditions of the
Proposition~\ref{prop:linext}, and let $<_L$ denote the corresponding
total order.

Assume first that $C_*$ is bounded; without loss of generality, we can
assume that $C_i=0$ for $i<0$, and $i>N$. Let $m=|M|$ denote the size
of the matching, and let $l=|\om|-2m$ denote the number of critical
cells.

We shall now inductively construct a~sequence of bases
$\om^0,\om^1,\dots,\om^m$ of $C_*$. Finer, each basis will be divided
into three parts: $\cc(\Omega^k)=\{c_1^k,\dots,c_l^k\}$,
$\cd(\Omega^k)=\{a_1^k,\dots,a_m^k\}$, and
$\cu(\Omega^k)=\{b_1^k,\dots,b_m^k\}$, such that $a_i^k=d(b_i^k)$, for
all $i\in[m]$.

We start with $\om^0=\om$ and the initial condition $b_i^0<_L
b_{i+1}^0$, for all $i\in[m-1]$. Since the lower index of $\fk_-(-,-)$
and $w_-(-\succ-)$ will be clear from the arguments, we shall omit it
to make the formulae more compact.

When constructing the bases, we shall simultaneously prove by
induction the following statements:

\nin $(i)$ $C_*=C_*[k]\oplus\ca_1^k\oplus\dots\oplus\ca_k^k$, 
where $C_*[k]$ is the subcomplex of $C_*$ generated by
$\Omega^k\sm\{a^k_1,\dots,a^k_k,b^k_1,\dots,b^k_k\}$, and $\ca_i^k$ is
isomorphic to $\atom(\dim b_i^k)$, for $i\in[k]$;

\nin $(ii)$ for every $x^k\in\cu(\om^k)\cup\cc(\om^k)$, 
$y\in\cc(\om^k)$, we have $w(x^k\succ y^k)=\sum_p w(p)$, where the sum
is restricted to those alternating paths from $x^0$ to $y^0$ which
only use the pairs $(a_i^0,b_i^0)$, for $i\in [k]$.

Clearly, all of the statements are true for $k=0$. Assume $k\geq 1$.
 
\vskip5pt

\nin {\it Transformation of the basis $\Omega^{k-1}$ 
into the basis $\Omega^k$:} set $a_k^k:=\bo b_k^{k-1}$,
$b_k^k:=b_k^{k-1}$, and $x^k:=x^{k-1}-w(x^{k-1}\succ a_k^{k-1}) \cdot
b_k^{k-1}$, for all $x^{k-1}\in\Omega^{k-1}$, $x\neq a_k,b_k$.

\vskip5pt

First, we see that $\om^k$ is a~basis. Indeed, assume $b_k^{k-1}\in
C_n$. For $i\neq n, n-1$, we have $\om_i^k=\om_i^{k-1}$, hence, by
induction, it is a~basis. $\om_{n-1}^k$ is obtained from
$\om_{n-1}^{k-1}$ by adding a~linear combination of other basis
elements to the basis element $a_k^{k-1}$, hence $\om_{n-1}^k$ is
again a~basis. Finally, $\om_n^k$ is obtained from $\om_n^{k-1}$ by
subtracting multiples of the basis element $b_k^{k-1}$ from the other
basis elements, hence it is also a~basis.

Next, we investigate how the poset $P(C_*,\om^k)$ differs from
$P(C_*,\om^{k-1})$. If $x\neq b_k$, we have $w(x^k\succ a_k^k)= 
\fk(\bo x^k,a_k^k)=\fk(\bo x^k,a_k^{k-1})=\fk(\bo x^{k-1},a_k^{k-1})-
w(x^{k-1}\succ a_k^{k-1})\cdot\fk(\bo b_k^{k-1},a_k^{k-1})=0$, where
the second equality follows from the fact that $\om_{n-1}^k$ is
obtained from $\om_{n-1}^{k-1}$ by adding a~linear combination of
other basis elements to the basis element $a_k^{k-1}$, and the last
equality follows from $\fk(\bo b_k^{k-1},a_k^{k-1})=1$.

Furthermore, since $\om_n^k$ is obtained from $\om_n^{k-1}$ by
subtracting multiples of the basis element $b_k^{k-1}$ from the other
basis elements, we see that for $x\in\om^k_{n+1}$, $y\in\om^k_n$,
$y\neq b_k$, we have $w(x^k\succ y^k)=w(x^{k-1}\succ y^{k-1})$.
Additionally, since the differential of the chain complex squares
to~0, we have $0=\sum_{z^k\in\om^k_n}w(x^k\succ z^k)\cdot w(z^k\succ
a_k^k)= w(x^k\succ b_k^k)\cdot w(b_k^k\succ a_k^k)=w(x^k\succ b_k^k)$,
where the second equality follows from $w(z^k\succ a_k^k)=0$, for
$z\neq b_k$.

We can summarize our findings as follows: all weights in the poset
$P(C_*,\om^k)$ are the same as in $P(C_*,\om^{k-1})$, with the
following exceptions:

\nin 1) $w(x^k\succ b_k^k)=0$, and $w(b_k^k\succ x^k)=0$, for $x\neq a_k$;

\nin 2) $w(a^k_k\succ x^k)=0$, and $w(x^k\succ a_k^k)=0$, for $x\neq b_k$;

\nin 3) $w(x^k\succ y^k)=w(x^{k-1}\succ y^{k-1})-
w(x^{k-1}\succ a_k^{k-1})\cdot w(b_k^{k-1}\succ y^{k-1})$, for
$x\in\om^k_n$, $y\in\om^k_{n-1}$, $x\neq b_k$, $y\neq a_k$.

In particular, the statement $(i)$ is proved.  Furthermore, the
following fact $(*)$ can be seen by induction, using 1), 2), and 3):
if $w(x^k\succ y^k)\neq w(x^{k-1}\succ y^{k-1})$, then $b_k^0\geq_L
y^0$. Indeed, either $y\in\{a_k,b_k\}$, or $y$ is critical, or
$y=a_{\ti k}$, for $\ti k>k$, such that $w(b_k^{k-1}\succ y^{k-1})\neq
0$.  In the first two cases $b_k^0\geq_L y^0$ by the construction of
$L$, and the last case is impossible by induction, and again, by the
construction of $L$.

We have $w(b_j^k\succ a_j^k)=w(b_j^{k-1}\succ a_j^{k-1})$, for all
$j,k$. Indeed, this is clear for $j=k$. The case $j<k$ follows by
induction, and the case $j>k$ is a consequence of the fact $(*)$.

Next, we see that the partial matching
$\cm^k:=\{(a_i^k,b_i^k)\,|\,i\in[m]\}$ is acyclic. For $j\leq k$, the
poset elements $b_j^k,a_j^k$ are incomparable with the rest, hence
they cannot be a~part of a~cycle. For $i>k$, we have $w(b_j^k\succ
a_i^k)= w(b_j^{k-1}\succ a_i^{k-1})$, by the fact~$(*)$. Hence, by
induction, no cycle can be formed by these elements either.

Finally, we trace the boundary operator. Let
$x^k\in\cu(\om^k)\cup\cc(\om^k)$, $y\in\cc(\om^k)$. For $x=b_k$ the
statement is clear. If $x\neq b_k$, we have $w(x^k\succ
y^k)=w(x^{k-1}\succ y^{k-1})- w(x^{k-1}\succ
a_k^{k-1})w(b_k^{k-1}\succ y^{k-1})$. By induction, the first term is
counting the contribution of all the alternating paths from $x^0$ to
$y^0$ which do not use the edges $b_l^0\succ a_l^0$, for $l\geq
k$. The second term contains the additional contribution of the
alternating paths from $x^0$ to $y^0$ which use the edge $b_k^0\succ
a_k^0$. Observe, that if this edge occurs then, by the construction of
$L$, it must be the second edge of the path (counting from $x^0$),
and, by the fact~$(*)$, we have $w(x^{k-1}\succ a_k^{k-1})=w(x^0\succ
a_k^0)$.  This proves the statement $(ii)$, and therefore concludes
the proof of the finite case.

It is now easy to deal with the infinite case, since the basis
stabilizes as we proceed through the dimensions, so we may take the 
union of the stable parts as the new basis for~$C_*$.
\qed

\begin{remark}
Even if the chain complex $C^*$ is infinite in both directions, one
can still define the notion of the acyclic matching and of the Morse
complex. Since each particular homology group is determined by
a~finite excerpt from $C^*$, we may still conclude that
$H_*(C_*)=H_*(C_*^\cm)$.
\end{remark}

\end{document}